\documentclass[final]{tlp}
\usepackage{hyperref}
\usepackage{graphicx}
\usepackage{aopmath}

\begin{document}
\bibliographystyle{./acmtrans}
\long\def\comment#1{}

\title{EPspectra: A Formal Toolkit for Developing DSP Software Applications}

\author[Hahnsang Kim, Thierry Turletti and Amar Bouali]{
Hahnsang Kim, Thierry Turletti\\
INRIA\\
Plan\`{e}te Project, 2004, Route des Lucioles BP93\\
06902, Cedex France\\
\and
Amar Bouali\\
Esterel-Technologies\\
885, av. Julien Lef\`{e}bvre\\
06270, Villeneuve, France\\
}

\pagerange{\pageref{firstpage}--\pageref{lastpage}}

\jdate{March 2002} \pagerange{35--48} \volume{\textbf{10} (3):}
\pubyear{2005} \submitted{March 2002}\revised{August
2004}\accepted{January 2005}

\setcounter{page}{1}

\maketitle

\label{firstpage}

\begin{abstract}
The software approach to developing Digital Signal Processing (DSP)
applications brings some great features such as flexibility,
re-usability of resources and easy upgrading of applications.
However, it requires long and tedious tests and verification phases
because of the increasing complexity of the software. This implies
the need of a software programming environment capable of putting
together DSP modules and providing facilities to debug, verify and
validate the code. The objective of the work is to provide such
facilities as simulation and verification for developing DSP
software applications. This led us to develop an extension toolkit,
{\sc EPspectra}, built upon {\sc Pspectra}, one of the first
toolkits available to design basic software radio applications on
standard PC workstations.

In this paper, we first present {\sc EPspectra}, an {\sc
Esterel}-based extension of {\sc Pspectra} that makes the design and
implementation of portable DSP applications easier. It allows
drastic reduction of testing and verification time while requiring
relatively little expertise in formal verification methods. Second,
we demonstrate the use of {\sc EPspectra}, taking as an example the
radio interface part of a GSM base station. We also present the
verification procedures for the three safety properties of the
implementation programs which have complex control-paths. These have
to obey strict scheduling rules. In addition, {\sc EPspectra}
achieves the verification of the targeted application since the same
model is used for the executable code generation and for the formal
verification.
\end{abstract}

\section{Introduction}
{\sc Esterel}~\cite{berry:96} is a synchronous programming language
targeted at reactive systems. {\sc Esterel} programs perform an
input-driven computation: wait for inputs and compute corresponding
outputs in a cyclic manner, referred to as a reaction. {\sc Esterel}
is also a formal language. The {\sc Esterel} system allows one to
provide specification, simulation, automatic code generation, and
verification. Taking into account that most of traditional
verification methods are concerned with proving properties only of
abstracted models of programs rather than programs themselves, the
{\sc Esterel} methodology allows one to directly verify the actual
code of {\sc Esterel} programs that corresponds to the targeted
implementation. It guarantees that the {\sc Esterel} programs
satisfy the properties to be proved on condition that all source
code is correctly compiled to the targeted code.

It still holds true that the number of transistors per integrated
circuit roughly doubles every 18 months according to Moore's
law\footnote{See {\tt
http://www.intel.com/research/silicon/mooreslaw.htm}}. Thus,
programming environments for Digital Signal Processing (DSP)
applications may no longer be required to rely on specialized DSP
hardware since the performance of general purpose processors and
peripheral equipment increases along with the high-tech curves. This
leads to the shift of hardware-operation functions into software. A
software approach to developing DSP applications allows the
following advantages: re-usability of existing hardware, ease of
upgrades, and more flexible applications. Nevertheless, it makes the
implementation of software applications more complex because of the
need for multi-disciplinary  knowledge of software architecture,
signal processing, real-time scheduling, networking protocols,
validation, etc. Furthermore, it requires an appropriate development
environment accessible to programmers.

The goal of this work is to develop a methodology to make the
implementation of DSP software applications easier by allowing the
code for specification, simulation and verification to be
executable. We make the best of the characteristics of {\sc
Esterel}, a formal as well as programming language.

We have developed an {\sc Esterel}-based extension toolkit, {\sc
EPspectra} built upon {\sc Pspectra}\footnote{It provides a signal
processing programming environment to implement portable DSP
applications on general-purpose workstations. See {\tt
http://www.sds.lcs.mit.edu/SpectrumWare/}}~\cite{bose:99,brett:00}.
In the {\sc EPspectra} system, the control part of DSP application,
which is to be verified eventually, is specified in {\sc Esterel}
and the data part, which contains DSP computation intensive modules,
is specified in C/C++. The behaviors of the control part are checked
out in simulation with {\sc Xes}~\cite{berry:99:xes} and its safety
properties are verified with {\sc Xeve}~\cite{bouali:98:xeve}. We
demonstrate the verification and the implementation of an example of
DSP software applications, the radio interface part of a GSM Base
Transceiver Station using {\sc EPspectra}. We also report the
results of performance comparison between the {\sc Esterel} based
implementation and the generic method based one.

This paper is structured as follows: Section~\ref{pandep} describes
the {\sc Pspectra} software architecture, which is divided into a
data part and a control part. It also describes an extension
toolkit, {\sc EPspectra}, of which the control part is re-designed
and implemented in {\sc Esterel}. Section~\ref{strl-method} presents
the features that are derived from the {\sc Esterel} methodology and
Section~\ref{schedulingtech} focuses on scheduling techniques
considering two models: the Data-Pull Model and Data-Reactive Model.
Once the extension toolkit has been described, we present in
Section~\ref{example} the implementation of a practical example
which corresponds to the radio interface part of a GSM Base
Transceiver Station. Three safety properties of the implementation
are verified in Section~\ref{verification}. In
Section~\ref{perfresults}, the performance results between {\sc
EPspectra} and {\sc Pspectra} are compared in terms of the capacity
of computation and the number of lines of code. Section~\ref{rel}
discusses the related work and the last Section concludes the paper
and presents future directions.

\section{Pspectra \& EPspectra}\label{pandep}
{\sc Pspectra}, developed by the SpectrumWare project at MIT, is a
real-time signal processing programming environment used to
implement portable DSP applications such as software radios on
general-purpose workstations. This environment includes a library of
portable (across platforms) DSP functions and an I/O subsystem. With
{\sc Pspectra}, the hardware part is minimal and the boundary
between software and hardware is shifted right up to the A/D
converter. This increases flexibility by bringing more functions
under software control.

The {\sc Pspectra} software architecture is partitioned into a
control part ({\it out-of-band components}) and a data part ({\it
in-band components}). This partitioning allows for a maximal re-use
of the computationally intensive DSP modules. The data part takes
care of the temporally sensitive and computationally intensive work,
while the control part deals with all code relating to scheduling
processing modules.

\subsection{Data Part}
The data part contains the code required to perform specific signal
processing tasks, access functions used by the control part to
configure and monitor the DSP tasks, and I/O functions that read
data from and write data into buffer. The data part consists of two
components: DSP modules and connectors. The DSP modules perform the
signal processing tasks and communicate with the control part via
the access functions. A connector can be thought of as a wire that
carries signals from the output of a processing module to the input
of the following processing module. The DSP modules are classified
as follows:
\begin{itemize}
\item {\it Sources} are specialized modules that have one or more
output ports and no input ports.
\item {\it Sinks} are specialized modules that have one or
more input ports and no output ports.
\item Intermediate modules have one or more input
ports and one or more output ports.
\end{itemize}
Each port must be connected to exactly one connector. Each signal
processing path has at least one source beginning computation and at
least one sink ending it.

\subsection{Control Part}
The control part is responsible for the creation of topology, the
modification of current data flow according to the system needs, the
control of the communications between DSP modules, the handling of
user interaction, and the monitoring of the data computation on each
DSP module. The data manipulated by the DSP modules flow from
sources to sinks. A DSP module reads input sample data from the DSP
module(s) directly preceding it, and performs some computation on
it.

\begin{figure}[htbp]
\centerline{\includegraphics[width=7cm]{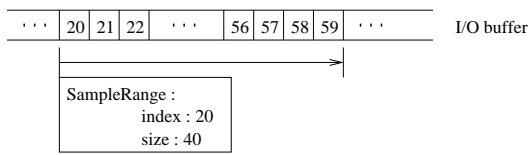}}
\caption{SampleRange: each data block is referenced with an index and a size}
\label{srange.pdf}
\end{figure}

To refer to the input and output data in the buffer, a parameter
called {\it SampleRange} is used in the DSP modules. This parameter
keeps track of a position of the data that each DSP module accesses.
As shown in Figure \ref{srange.pdf}, a SampleRange contains two
pieces of information: an {\it index} identifying a starting point
from which to read data into the buffer and a {\it size} identifying
the amount of data to be read.

All DSP modules include an {\it estimating function} and a {\it
computing function}. The estimating functions in DSP modules specify
a SampleRange used by computing functions with reference to the
SampleRange parameter of the preceding modules and inform the
following modules of their SampleRange parameter. In addition,
estimating functions have to ensure that the same data is not
computed more than once. Computing functions start when estimating
functions successfully return, and they manipulate the data that
estimating functions have scheduled.

\subsection{Esterel-based Architecture}
Even though {\sc Pspectra} provides features such as dynamic
flexibility, portability, and re-usability for software
implementations, it lacks the functionality of simulation, testing,
and formal models accessible to developers. Data-intensive
activities and control-driven handling activities require different
programming techniques.

\begin{figure}[htbp]
\centerline{\includegraphics[width=4.5cm]{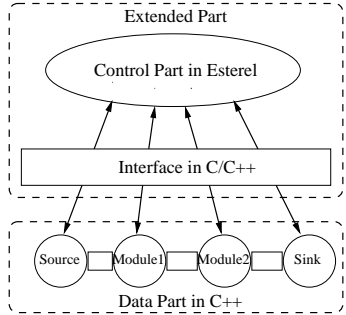}}
\caption{Architecture of an Esterel-extension}
\label{arch.pdf}
\end{figure}

In an {\sc Esterel}-based approach, as shown in
Figure~\ref{arch.pdf}, the architecture is composed of an extended
part and the data part on the whole. The extended part is
partitioned into the control part in {\sc Esterel} and the interface
part in C/C++. In the control part, the components of DSP modules
are instantiated, initialized and scheduled. The interface part is
represented as an interface to link the {\sc Esterel}-written
control part to the C++-written data part. The data part in C++ is
where DSP algorithms are run.

\begin{figure}[htbp]
\centerline{\includegraphics[width=9cm]{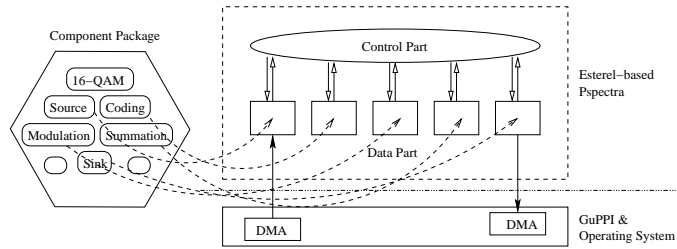}}
\caption{The Esterel-based Pspectra environment}
\label{env.pdf}
\end{figure}

As a whole, as shown in Figure~\ref{env.pdf}, the {\sc
Esterel}-based {\sc Pspectra} software environment ({\sc EPspectra})
contains the following: the component package that provides a
library of computational functions for the data part and the General
Purpose PCI Interface (GuPPI\footnote{See {\tt
http://www.sds.lcs.mit.edu/SpectrumWare/guppi.thml}}). It allows the
sampled signal data to be directly transferred in and out of memory
of the workstation via Direct Memory Access (DMA).

\subsection{Esterel Implementation of the Control Part}
\label{implcontrol}
\begin{figure*}[htbp]
\centerline{\includegraphics[width=14cm]{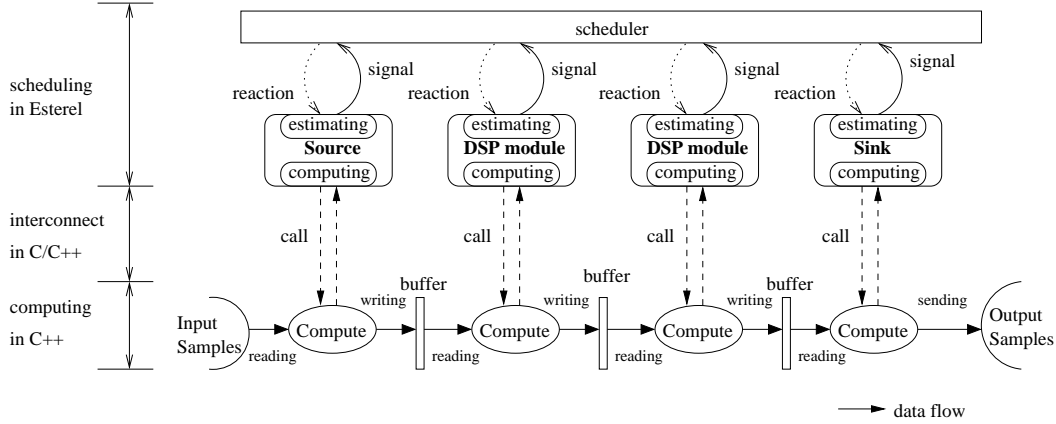}}
\caption{Architecture of the control part of EPspectra}
\label{epspec.pdf}
\end{figure*}

Figure~\ref{epspec.pdf} shows the architecture of {\sc EPspectra} in
more detail. In all the modules, the computing functions follow the
estimating functions. A scheduler first triggers the estimating
function on the source by sending a control signal. When the
estimating function is completely performed, the source emits an
ack-signal to the scheduler that will allow it to perform the
estimating function on the next module. At the same time, the
computing function on the source is performed to compute the sample
data. Afterwards, the source is required to wait until the
ack-signal coming from the next module is received. As soon as the
next module is ready to compute the corresponding sample data, the
source repeats the same procedure to manipulate the continuous
sample data.

When each intermediate module gets a control signal from its
preceding module(s) via the scheduler, it starts computation and
then transmits the computed sample data to the next modules while it
sends an ack-signal to the preceding modules. The sinks perform the
same operation as intermediate modules except that there is no next
module.

\section{The Esterel Methodology}\label{strl-method}
{\sc Esterel} belongs to the family of synchronous reactive
languages, such as {\sc Lustre}~\cite{hal:92}, {\sc
Signal}~\cite{ben:91:signal} and {\sc StateCharts}  \cite{harel:87}.
{\sc Esterel} provides powerful constructs to express sequencing,
parallel behavior, and preemption. It also provides a communication
mechanism with signal broadcasting. These constructs are
particularly suited for the programming of a reactive system
containing the control-dominated part. The {\sc Esterel} language
has clean mathematical semantics that interpret an {\sc Esterel}
program as a Finite State Machine (FSM), a state-graph model with
labels over the graph edges. The FSM model represents exhaustively
all the possible states that the program can be in and all the
behaviors that the program can perform between the states. The main
features brought by the {\sc Esterel} methodology are:
\begin{itemize}
\item {\bf Specification}: Although {\sc Esterel} is relatively
simple, it is expressive and concise enough to program complex
controllers.
\item {\bf Simulation}: The {\sc Esterel} system provides symbolic
debugging simulation with the symbolic debugging simulator {\sc
Xes}. The simulation environment is based on the Finite State
Machine (FSM) model. The simulator is coupled with the formal
verification environment.
\item {\bf Automatic code generation}: The {\sc Esterel} system
compiles an {\sc Esterel} program into an executable C program with
a C interface that is easy to connect with hand-written C code. The
C code represents the FSM model exactly.
\item {\bf Formal verification}: The FSM model allows one to
perform model-checking to verify its properties. When any property
is not satisfied, the verifier generates the corresponding
counter-example input-sequence. This counter-example can be played
back using {\sc Xes}. More details of model-checking are given in
Section~\ref{verification}.
\end{itemize}
Hence, {\sc Esterel} is not only a programming language, but also
provides a formal method, which means there is no gap between
specification or simulation and execution. Using the {\sc Esterel}
methodology, the procedure verifying the properties of an {\sc
Esterel} program is the following:
\begin{itemize}
\item[i.] describe the properties satisfying the correctness of
an {\sc Esterel} program,
\item[ii.] compile the {\sc Esterel} program in parallel with
{\it observer}, the program that describes properties and
\item[iii.] check for satisfaction or violation of the properties
using the {\sc Esterel} model-checker {\sc Xeve}.
\end{itemize}

\section{Scheduling Techniques}\label{schedulingtech}
It is useful to review existing definitions of real-time systems
before describing the statistical real-time model. Although there
are many different definitions of real-time constraints in the
literature, we can generally classify them into {\it hard} real-time
and {\it soft} real-time constraints~\cite{jensen:94}. In hard
real-time systems, the overall time consumption of DSP modules is
strictly limited. In other words, all the time critical functions
have deadlines which must always be met in order for the system to
function properly. This domain includes safety-critical real-time
applications such as space rockets, aircraft automatic pilots, air
traffic control, car vital systems, and some medical equipment. On
the other hand, soft real-time systems are not well defined. They
are generally thought of as real-time systems that can still
function reasonably well even if deadlines are occasionally missed.
Indeed, the reliability of a system relies on the accuracy of the
estimates.

{\sc EPspectra} and {\sc Pspectra} run on general purpose
workstations in an operating system (Linux OS) without {\it
explicit} real-time support. Instead, by taking advantage of the
ability to sometimes process data faster than in real-time, jitter
in the computation time of some functions can be absorbed. This
provides a real-time scheduling mechanism for dealing with frequent
small-scale time variability. Resource unpredictability may result
in the processing time occasionally exceeding the real-time rate,
but the average processing rate can still be well below the
real-time threshold. Thus, there is a trade-off between higher
average throughput and jitter in the computation time. In order to
deal with the larger variations, the concept of {\it statistical
real-time performance} is introduced with the following
characteristics:
\begin{itemize}
\item the cumulative distribution of the number of cycles required
to complete a task,
\item a desired real-time bound and
\item a specification of the action that must be performed when
the deadline is not met.
\end{itemize}
This is a kind of soft real-time constraint since deadlines can be
missed without disastrous consequences. The probability that the
task will be completed within the desired time bound can be
expressed from the cumulative distribution of cycles required to a
given application. This is possible since the statistics associated
with the execution time are consistent. Note that if the task is
completed with a probability of one, then the system will provide
hard real-time constraints.

Different actions are possible when a deadline is missed. For
example, one can abort computation and drop the remaining data,
replace the remaining data by a special value or partially estimated
data from the result, or start processing the next slice of data
while the current processing  job continues in parallel.

Instead of extending the real-time paradigm across the whole system,
{\sc EPspectra} and {\sc Pspectra} extend the boundaries of the
virtual time environment by (i) time-stamping and temporally
decoupling sampled information at the edge of the system and (ii)
providing a virtual time programming environment in which it is
possible to implement applications that process temporally sensitive
information.

\subsection{DPM: Data-Pull Model}\label{DPM}
Let us account for the Data-Pull Model (DPM) on which the control
part of {\sc Pspectra} is based before looking into the
Data-Reactive Model (DRM). The DPM relies on a ``lazy evaluation
approach''~\cite{johnsson:84}. Lazy evaluation so-called call by
need has been proposed as a method for executing functional
programs. The advantages of using the DPM in {\sc Pspectra} include:
improved computational efficiency resulting from the benefits of
lazy evaluation, the rapid response to changes in the processing
requirements, and the caching benefits with a good locality of data
reference by means of lazy evaluation. Further details concerning
these advantages are described in~\cite{bose:99}.

{\sc Pspectra} performs parallel processing for data computation of
DSP modules, generating multiple threads. However, the overhead of
synchronization between threads which share the same data may
degrade the performance of parallel processing. Suppose that there
is an application composed of two sources, two sinks, and several
intermediate modules. The two independent sinks are connected to the
same intermediate module. According to the DPM, the two sinks are
only processed alternately. In addition, when the sequential
processing chain is created, sample data is processed by passing it
through this chain, but the next sample data is not processed before
the process of the sample data is completed. More specifically, it
is not possible to interleave the computation chain of the current
sample data and that of the next sample data.

\subsection{DRM: Data-Reactive Model}
\begin{figure}[htbp]
\centerline{\includegraphics[width=7cm]{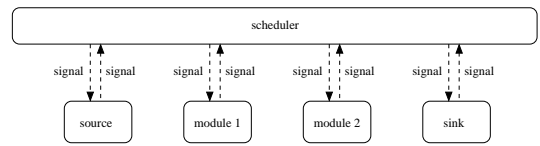}}
\caption{Data-Reactive Model}
\label{drm.pdf}
\end{figure}

In contrast to the DPM, the DRM makes use of a software pipelining
method~\cite{allan:95}. It allows the reduction of the idle time
between the beginning and the end of computation operations. It
accelerates computation operations as well as computation-intensive
scheduling. Figure \ref{drm.pdf} shows the architecture of the DRM
specified in {\sc Esterel}. All the modules wait for input signals
and compute the corresponding output signals. The DRM allows the
benefit of the well-formed semantic properties of {\sc Esterel} such
as parallel composition and hierarchical automata, introduced in
\cite{berry:96}.

The data processed on the source is pushed into the sink through the
operations of the intermediate modules. Since all DSP modules react
on available data, a scheduler determines the relation among DSP
modules, and activates or deactivate them according to the relation.
The scheduling approach is the following:
\begin{itemize}
\item data computation starts on the source,
\item whenever data on DSP modules are available, they start computing it, and
\item the corresponding data is consumed on the sink.
\end{itemize}
The scheduler monitors and controls the communications of DSP
modules. As soon as the sources finish computing the data, they emit
certain signal(s) triggering the computation of the corresponding
data on the following modules and then wait for ack-signals from
them. The DSP modules wait for two events: available data from the
preceding modules and ack-signals from the following modules. Here,
receiving ack-signals implies the completion of computation of the
previous data. When receiving both of them, the DSP modules compute
the available data, and then convey the computed data to the
following modules. At the same time, they emit ack-signals to the
preceding modules simultaneously. The corresponding data are finally
consumed on the sinks. The DRM has two features of scheduling: a
software pipelining scheduling method and the data dependencies.

\subsubsection{Software pipelining schedule}\label{features}
\begin{figure}[htbp]
\centerline{\includegraphics[width=7cm]{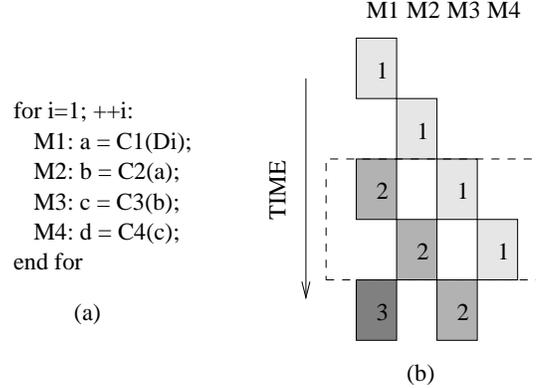}}
\caption{(a) Loop body code. (b) Execution schedule of iterations.}
\label{pipeline}
\end{figure}

The software pipelining scheduling method makes use of parallel
processing among DSP modules {\it at the operation-scheduling
level}, not at the instruction level. Let us look at the loop body
of Figure~\ref{pipeline}(a). Each set {\bf M}\footnote{Note that M
represents a set of operations of each module, not an operation
itself.} of an iteration depends on the previous set of operations
as well as the previous iteration. As shown by the execution
schedule of Figure~\ref{pipeline}(b), the set of operations of the
2nd iteration of {\bf M1} depends on, and must follow the set of
operations of the 1st iteration of {\bf M2}. From this basic
software pipelining scheduling method, speed-up of the execution
rate can be expected.

\subsubsection{Data Dependencies}
DSP modules of a DSP application based on the DRM are {\it dependent
on data} associated with its topology. A {\it
dependence}~\cite{allan:95} exists between two operations if
interchanging their order affects the results. Dependencies
constrain what can be done in parallel. Let $O_1$ and $O_2$ be
operations such that $O_1$ precedes $O_2$. $O_2$ must follow $O_1$
if $O_2$ reads data written by $O_1$. $O_2$ is said to be {\it data
dependent} on $O_1$. {\it Data} dependence between two operations is
extended to data dependence between two operational modules. There
is another reason that one operation must wait for another
operation. A {\it control} dependence exists between $S_1$ and $S_2$
if the execution of statement $S_1$ determines whether or not
statement $S_2$ is executed. Therefore, even though $S_2$ is able to
execute because of the available data, it may not execute because it
is not known whether it is needed.

\begin{figure}[htbp]
\centerline{\includegraphics[width=8.5cm]{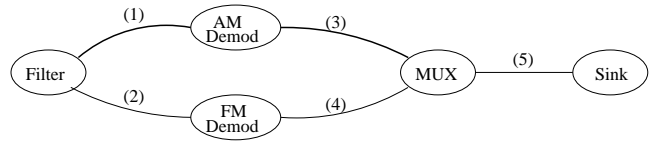}}
\caption{A diagram showing dependencies}
\label{dependence}
\end{figure}

The DRM considers data dependencies, not control dependencies.
Figure~\ref{dependence} gives an example of this. It shows part of
an audio application that switches between Amplitude Modulation (AM)
and Frequency Modulation (FM) demodulators, consisting of the
filter, AM demodulation, FM demodulation,multiplex and sink modules.
The audio application has data dependencies represented as (1), (2),
(3), (4), and (5) and all the statements pertaining to the execution
of modules. The control program is required to change the execution
topology with the establishment of either (1) and (3), or (2) and
(4) after the Channel Filter operation. Thus, it is necessary to
have control dependencies as well as data dependencies between the
Channel Filter and the AM demodulator, or between the Channel Filter
and the FM demodulator. It implies the need of the {\it dynamic}
reconfiguration that enables the execution topology to be adapted to
the changeable environment.

\section{An example of Application: The Radio Interface Part of a GSM BTS}
\label{example} As an example of implementation using {\sc
EPspectra}, this section describes the general architecture of a GSM
network and the radio interface part of a GSM base station. The
following sections describe the verification procedure of the three
safety properties of the implementation that should be satisfied. In
addition to the verification, the performance comparison between
automatically generated code programs and hand-written code programs
is analyzed.

\begin{figure*}[htbp]
\centerline{\includegraphics[width=13cm]{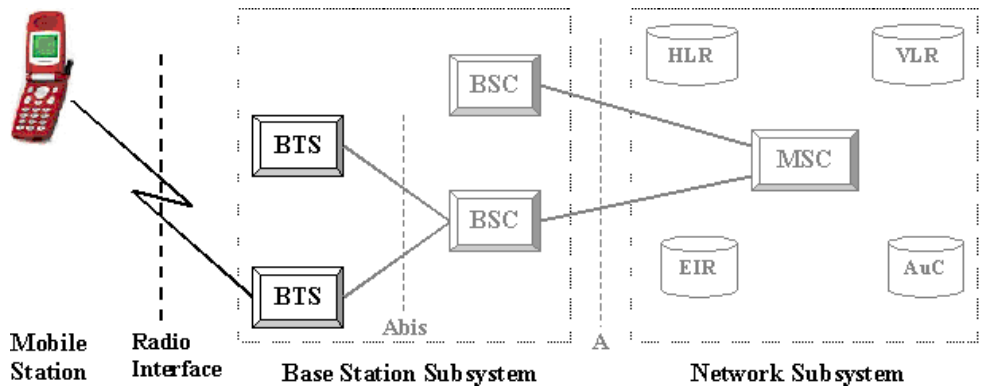}}
\caption{Architecture of GSM Network}
\label{gsmnetwork.pdf}
\end{figure*}

As shown in Figure~\ref{gsmnetwork.pdf}, the GSM network can be
generally divided into three main parts: the Mobile Station (MS),
the Base Station Subsystem (BSS), and the Network SubSystem (NSS).
The MS is the physical equipment used by a subscriber, most often a
normal hand-held cellular telephone. The BSS connects the MS and the
NSS. It is in charge of transmission and reception. The BSS consists
of a Base Transceiver Station (BTS) and a Base Station Controller
(BSC). The BTS comprises radio transmission and reception devices
and also manages the signal processing related to the air interface.
Each BTS has one to sixteen transceivers, depending on the density
of users in the cell. The BSC controls a group of BTS and manages
its radio resources, mainly through the allocation, release and
hand-over of radio channels. The Mobile Switching Center (MSC) is
the central component of the NSS. It performs the switching
functions of the network and also provides connection to other
networks. In addition, there are several kinds of registers, namely
the Home Location Register (HLR), the Visitor Location Register
(VLR), Equipment Identity Register (EIR), and the Authentication
Center (AuC). The further description of the GSM system is given
in~\cite{mouly:93}.

We focus on the implementation of the GSM radio interface part
between the MS and the BTS, particularly on the BTS side. It
provides a multiple-access scheme and operations for the
transformations between source information and radio waves. The
implementation of the multiple access scheme has been excluded from
our work. Instead, we present and implement the operations that have
to be performed to pass from the speech source to radio waves and
vice-versa.

\subsection{Sequence of Operations between source information and radio waves}
\begin{figure}[htbp]
\centerline{\includegraphics[width=5cm]{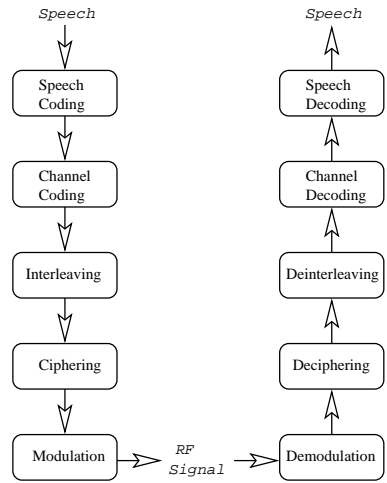}}
\caption{Sequence of operations from speech to radio waves and back to speech}
\label{sequenceofgsm.pdf}
\end{figure}

The sequence of operations for the radio interface of a GSM BTS is
shown in Figure~\ref{sequenceofgsm.pdf}. Basically, after having
transformed speech into compressed data blocks in speech coding,
channel coding adds redundancy to the data blocks. The data blocks
are interleaved and spread into pieces in interleaving, which are
combined with flags to build up the bursts. Ciphering is applied to
these bursts and then the resulting data is used to modulate the
carriers in modulation. The reverse transformations are performed on
the other side.

\begin{itemize}
\item {\bf Speech coding} algorithm, Regular Pulse Excitation with Long Term
Prediction (RPE/LTP)~\cite{94:european} produces data blocks of 260
bits every 20ms.
\item {\bf Channel coding} introduces redundancy into
the data flow, increasing its rate by adding information calculated
from the source data, in order to allow the detection or the
correction of signal errors introduced during transmission. It forms
a complete coded speech frame of 456 bits.
\item {\bf Interleaving} consists in mixing up the bits of
several code words, which in the modulated signal are spread over
several code words. GSM coding blocks are interleaved on 8 bursts
each of which consists of 57 bits.
\item {\bf Ciphering} performs an exclusive or (XOR) operation
between 2 bursts of each block and a secret recipe known only by the
mobile station and BTS.
\item {\bf Modulation} transforms the binary signal into
a Gaussian Minimum Shift Keying (GMSK)~\cite{murota:81:gmsk}.
\item Once radio waves are captured by the antenna, the portion
of the received signal which is of interest to the other side
corresponding to radio waves to source information is determined by
the multiple access rules. {\bf Demodulation} takes place in this
portion.
\item {\bf Deciphering} performs the same operations by reversing
the ciphering algorithm.
\item {\bf Deinterleaving} merges two different 8-burst blocks
into a 456-bit code word.
\item {\bf Channel decoding} involves reconstructing the source
information from the output of the demodulator, using the added
redundancy to detect or correct possible errors in the output from
the demodulator.
\item {\bf Speech decoding} reconstructs the speech by passing
the residual pulse first through the long-term prediction filter,
and then through the short-term predictor.
\end{itemize}

\section{Formal Verification}\label{verification}
In this section, we explain what formal verification of {\sc
Esterel} programs means. As mentioned, {\sc Esterel} is both a
formal modeling language and a programming language. {\sc Esterel}
benefits from clear mathematical semantics that characterize a
program as an FSM model. The {\sc Esterel} FSM model is defined as a
structure $(I,O,S,s_0,T)$ where $I$ is a set of input signals, $O$ a
set of output signals, $S$ a set of states, $s_0$ the initial state,
and $T$ a transition relation. $T$ is a set of 4-tuple
$(s,i,o,s^\prime)$, which represents a transition from $s$ to
$s^\prime$ whenever the input event $i$ is true, generating the
output event $o$. The FSM model is the one that is used for
simulation, execution and formal verification. A state of the FSM
model is a stable configuration of the control points of the
program. A transition from a system state is a reaction to some
input event: the reaction leads to a new stable system. The FSM
model has the advantage of exhaustively exhibiting the program
behaviors. Formal verification is the activity of proving properties
of programs and systems in a mathematical sense. In other words,
verification consists in verifying the satisfaction of a set of
properties over a FSM model of the program or system behavior.

Generally speaking, there are two types of properties that can be
expressed: safety properties~\cite{alpern:86} and liveness
properties~\cite{alpern:85}. Safety properties express the fact that
``something bad will never happen.'' Liveness properties express the
fact that ``something good will eventually happen.'' For example, a
typical safety property is "The elevator will never move while the
door is open" and a typical liveness property is "If someone calls
the elevator, then the elevator will eventually come". In our
experience, most of the properties are safety ones. When liveness is
concerned, it is often reducible to {\it bounded} liveness, which is
fundamentally a particular form of safety properties. Bounded
liveness properties express the fact that ``something good will
eventually happen in at most k times units,'' where k is a constant.
For example, we can transform the liveness property of the elevator
into a bounded liveness as follows: "If someone calls the elevator,
then the elevator will eventually come in less than 5 minutes". Let
us look into a way to directly apply these properties to the {\sc
Esterel} system.

\subsection{Observer Properties}
In the {\sc Esterel} system, the users directly express the
properties using the {\sc Esterel} language. Let us consider a
simple property that requires the following condition: ``At each
state, if signal A and B are present, then signal C in the next
state should be present unless signal R is present. Otherwise it
falls into an error state''. In {\sc Esterel}, this property is
written as follows:

\begin{verbatim}
module OBSERVER:
input A, B, C, R;
output BUG;
loop
   present [A and B] then
      pause;
      abort
         present C else emit BUG end
      when R
   else
      pause
   end present
end loop
end module
\end{verbatim}
\noindent The {\tt pause} statement waits for one time unit. The
{\tt abort ... when cond} construct kills its body as soon as the
condition {\tt cond} is true. This formal verification consists in
checking if the signal properties such as {\tt BUG} above can be
emitted in some reachable states and for some input events. If the
property is violated, the {\sc Xeve} model-checker generates an
input sequence of events that would have produced the error state.

\subsection{Xeve}
{\sc Xeve} takes as inputs the FSM model expressed as a set of
boolean equations in {\sc Blif} format generated by the {\sc
Esterel} compiler. It makes use of the symbolic state space
construction algorithm by means of Binary Decision Diagrams
(BDDs)~\cite{bryant:86}, the internal representation of an FSM model
for the reachable state space. {\sc Xeve} provides two verification
functions: minimising the number of states of the FSM model and
checking the emission status of output signals. The first function
is performed with respect to an equivalence notion called {\it
symbolic bisimulation}~\cite{simone:93}. The second function checks
two states for output signals: {\it possibly emitted}, which means
there exists a reachable configuration that some inputs lead to the
emitted output signals, and {\it never emitted}, which means there
exists no reachable configuration that some inputs lead to the
emitted output signals. More details on {\sc Xeve}'s verification
technique can be found in~\cite{bouali:98:xeve}.

\subsection{Properties of the GSM Programs}\label{TempFp}
Basically, all processing modules do their behaviors in parallel.
Let us take a look at the following example.

\begin{figure}[htbp]
\centerline{\includegraphics[width=8.5cm]{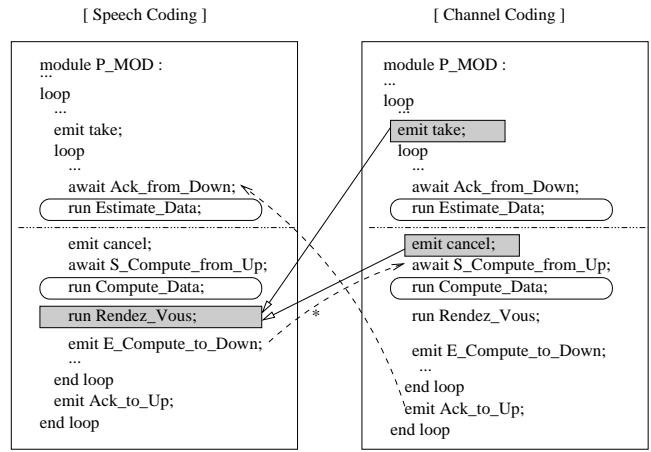}}
\caption{Signal passing diagram between two adjacent modules}
\label{rendez-vous.pdf}
\end{figure}

\begin{verbatim}
module GSMsource2wave
   ...
   run source/SOURCE
    ||
   run speechcoding/P_MOD
    ||
   run channelcoding/P_MOD
    ||
   run interleaving/P_MOD
    ||
   run ciphering/P_MOD
    ||
   run modulation/P_MOD
    ||
   run sink/SINK
   ...
end module
\end{verbatim}
Unless carefully programmed, the process of a module may prevent the
process of the other modules from running due to missing signals.
Let us look at Figure~\ref{rendez-vous.pdf}. The performance of
parallelism can be enhanced as the process of an inside module is
partitioned into two parts (i.e. the estimating and the computing
function parts) running in parallel. It requires the cautious
synchronization between the process of a module and that of another.

In Figure~\ref{rendez-vous.pdf}, the `Ack\_from\_Down' signal of the
speech coding module is synchronized with the `Ack\_to\_Up' signal
emitted by the channel coding module. As soon as `Ack\_from\_Down'
is received, the estimating function is performed on the speech
coding module (i.e. run Estimate\_Data). The `S\_Compute\_from\_Up'
signal on the channel coding module is synchronized with the
`E\_Compute\_to\_Down' signal emitted in the speech coding module.
This synchronization activates the computing function on the channel
coding module (i.e. run Compute\_Data). However, Estimate\_Data and
Compute\_Data submodules contain {\sf loop} statements including
{\it ticks}\footnote{{\it Tick} introduced in {\sc Esterel} is
thought of as logical time which represents the activation clock of
a reactive program.}, the number of which being consumed is
determined at the run time execution. It may cause deadlock to
happen by the channel coding module to miss the signal coming from
the speech coding module (which corresponds to the asterisked arrow
($\ast$) in Figure~\ref{rendez-vous.pdf}). There are occasions when
the number of ticks being consumed are not transparent to
programmers. We, therefore, add an {\it explicitly} synchronizing
mechanism called {\it Rendez-Vous}. As shown in
Figure~\ref{rendez-vous.pdf}, the `take' signal on the channel
coding module validates Rendez-Vous submodule on the speech coding
module, which results in suspending the successive process of the
speech coding module. This delays in emitting the
`E\_Compute\_to\_Down' signal on the speech coding module.
Afterwards, on receiving the `cancel' signal coming from the channel
coding module, {\it Rendez-Vous} kills this suspension.

The GSM programs process sample data; the data processed on the
sources ends up being consumed on the sinks. All modules contain
{\sf loop} statements, which means that the programs may stall or
may be in a situation in which some critical stage of a task is
unable to finish. This fact must be verified for the safety of the
programs. Accordingly, the requirements that should be satisfied by
the above model are the following:
\begin{itemize}
\item[R1] The signal emitted by a module is always caught by
the opposite modules (referred to as {\it Rendez-Vous}).
\item[R2] The computed sample data on the source(s) will eventually
be consumed on the sinks.
\item[R3] Whenever the modules receive input signals,
they emit the corresponding output signals within a {\it bounded}
time-period.
\end{itemize}
Each safety property is then translated into an {\sc Esterel}
observer. The safety properties and the corresponding translations
are as follows:
\begin{itemize}
\item[S1] Deadlock freedom: an important safety property is
deadlock freedom. In the GSM program, deadlock occurs when one
misses signals that should be received. The {\it Rendez-Vous}
mechanism aims to avoid this synchronization deadlock by
establishing an explicit synchronization between at least two
signals of modules running in parallel. To guarantee that the
program will never deadlock, it is sufficient to verify the {\it
Rendez-Vous} mechanism, namely, by checking the satisfaction of the
following safety property: any state at which the module emits
`E\_Compute\_to\_Down' is preceded by a state at which the opposite
ones are ready to receive `S\_Compute\_from\_Up'. This is stated by
the following {\sc Esterel} observer:

\begin{verbatim}
module S1:
input ReadytoReceive, E_Compute_to_Down;
output S1_VIOLATED;
loop
   await E_Compute_to_Down;
   abort
      emit S1_VIOLATED
   when pre(ReadytoReceive);
end loop
end module
\end{verbatim}

\item[S2] Correctness: a major scheduling task of the GSM programs is
to correctly deliver certain sample data computed on a source up to
a sink by applying a sequence of operations to the corresponding
data. The procedure begins from the source receiving
`Ack\_from\_Down' and ends when the sink emits `Ack\_to\_Up'.
However, we note that all sample data computed on the source is not
always consumed by the sink in the end. In fact, a certain amount of
sample data can be skipped, depending on the specific conditions,
{\it e.g.}, a missed deadline happens since it is scheduled based on
soft real-time constraints, there are changes to a type of
modulation algorithm or an event, such as {\it reset}, occurs from
the outside environment.

Each module consumes one or two ticks for an iteration of the {\sf
loop} statement from the input sample data to the corresponding
output sample data. The GSM programs are divided into the operations
of the transmission from source to radio wave and back. Each of them
consists of five modules plus a source and a sink (See
Figure~\ref{sequenceofgsm.pdf}). Suppose that each module consumes
two ticks for a sample data, the sink finishes computing the sample
data in no more than fourteen ticks (= {\it D}). The correctness
property is as follows: for a state receiving `Ack\_from\_Down' on
the source, a state emitting `Ack\_to\_Up' on the sink follows no
more than {\it D} position. This is stated by the following {\sc
Esterel} observer:

\begin{verbatim}
module S2:
constant D;
input Ack_to_Up, Ack_from_Down;
output S2_VIOLATED;
   await Ack_from_Down;
   abort
      await D tick;
      emit S2_VIOLATED
   when Ack_to_Up;
end module
\end{verbatim}

\item[S3] Safety-liveness: every module behaves like
a sub-reactive program which waits for inputs and computes
corresponding outputs in a cyclic manner. Each module contains a
{\sf loop} statement with a certain condition to exit. There is
possibly a situation where some critical stage of a task is unable
to finish, referred to as {\it livelock}. If one module is
livelocked, the other modules would be blocked. The number of ticks
being consumed for a period of receiving and responding to inputs is
proportional to the length of a path between the module and the
sink.

Figure~\ref{opath} shows a signal-passing scenario of the GSM
program performing operations from source information to radio
waves. All the modules except the source and the sink wait for two
input signals from the previous and next modules: {\it await
AckfromUp\&Dwn}. In an initial state, the signal coming from the
next module is given as on. We consider quantifying the total number
of ticks being consumed to compute given sample data on the source
up to the sink. Each module contains two `pause' identified by
`await tick' and seven modules compose a sequence of operations.
Therefore, fourteen ticks are consumed in total, which is maximum
number because the courses of operations for one sample data and
another are interleaved.

\begin{figure*}[htbp]
\centerline{\includegraphics[width=13cm]{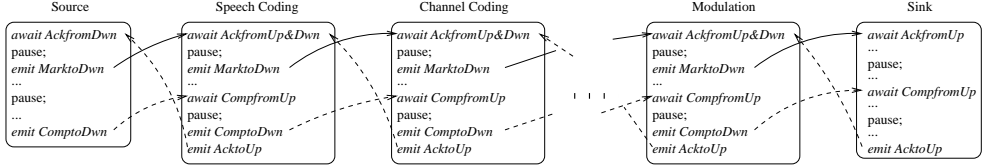}}
\caption{A signal-passing scenario of the GSM program corresponding
to from source to radio waves}
\label{opath}
\end{figure*}

Considering that each module consumes two ticks in an iteration of a
{\sf loop} statement, the source receives an ack-signal in no more
than fourteen ticks (={\it D}) and yet {\it D} is also proportional
to the length of the signal passing chain. The general form of the
property is as follows: if $I_s$ holds at position $j$, then $O_s$
holds at position $k$, for $j\leq k \leq j+D$. This is stated by the
following {\sc Esterel} observer:

\begin{verbatim}
module S3:
constant D;
input Is, Os;
output S3_VIOLATED;
loop
   await Is;
   abort
      await D tick;
      emit S3_VIOLATED
   when Os;
end loop
end module
\end{verbatim}
This property can be applied separately to the source, the sink and
the others. For example, ({\it AckfromDwn}, {\it ComptoDwn}) for the
source, ({\it AckfromUp}, {\it AcktoUp}) for the sink, and ({\it
AckfromUp\&Dwn}, {\it AcktoUp}) for intermediate modules are event
predicate pair $(I_s,O_s)$ being observed.
\end{itemize}

At the phase of combining the GSM programs with observers in the
properties verifying procedure, the following program is defined
consisting of the {\tt GSM} program to be verified and three
observers, {\tt S1}, {\tt S2} and {\tt S3} to verify. In {\sc Xeve},
the occurrence of {\tt S1\_VIOLATED}, {\tt S2\_VIOLATED}, and {\tt
S3\_VIOLATED} is checked. We note that the {\tt GSM} program is
compiled directly into an executable code without modification. We
will analyze the performance of the {\tt GSM} program in Section 7.

\begin{verbatim}
module VERIFY_PROGRAM:
constant D:=14 : integer;
input <the program inputs>;
output <the program outputs>,
       S1_VIOLATED,
       S2_VIOLATED,
       S3_VIOLATED;

   run GSM
    ||
   run S1
    ||
   run S2
    ||
   run S3
end module
\end{verbatim}

\subsection{Verification Process}
\begin{figure}[htbp]
\centerline{\includegraphics[width=8.5cm]{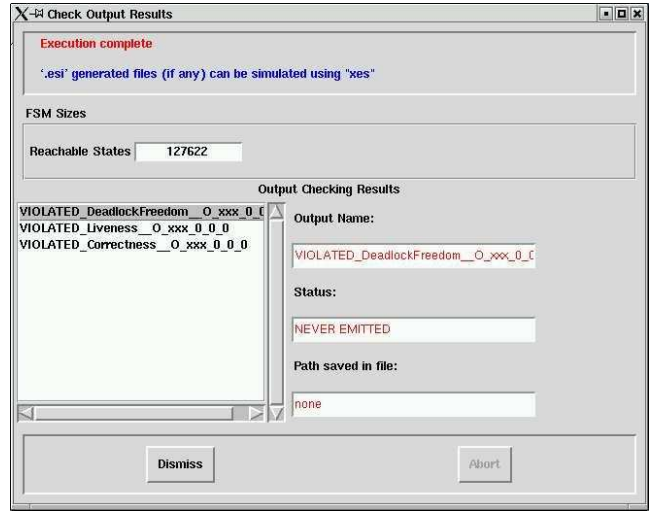}}
\caption{Screen-shot of a verification result of checking
the status of output signals}
\label{dnshot}
\end{figure}

We verified the satisfaction of the properties in the GSM programs
by confirming {\it never emitted} the property-signals including
{\sc violated\_deadlockfreedom}, {\sc violated\_correctness}, and
{\sc violated\_liveness} using {\sc Xeve}. Figure~\ref{dnshot} shows
the result of verifying the status of the property-signals of the
GSM program containing the operations of the transmission from
source to radio waves in {\sc Xeve}.

Generated reachable state space of two GSM {\sc Esterel} programs
(one describing the operations from source to radio waves and the
other describing those of backward) amounts to 127622 and 116972
states, respectively.

The number of nodes of the BDD graphs representing these reachable
state spaces are 66965 and 63390 respectively. It takes each program
about 273 and 236 seconds-CPU time on Linux machine with 600Mhz
Pentium processor and 516 RAM
 to generate the reachable state space.
Note that these amounts were generated by the combination of the
properties and the implementation of the {\sc Esterel} programs.

\section{Performance Results}\label{perfresults}
Our performance analysis in the verification process was carried out
on a Pentium 600MHz machine with 512MB of core memory and 516MB of
swap space on Linux kernel 2.2.15. {\sc Esterel} programs are
compiled with the version 6.03 of the {\sc Esterel} compiler into
{\sc Blif} formats\footnote{{\em Berkeley Logical Interchange
Format} is an ASCII format developed at the university of Berkeley
to describe a logic-level hierarchical circuit in textual form.} and
then optimized by {\sc Remlatch}~\cite{sentovich:96} and {\sc
Sis}~\cite{sentovich92}. The {\sc Remlatch} processor is used to
optimize the state encoding of the circuit and {\sc Sis} is used to
reduce the combinational logic introduced by the sequential
optimisation of {\sc Remlatch}. The optimized Blif code is
translated into standard C code by the {\sc Esterel} compiler. The
executable code\footnote{The executable code is obtained by gcc
version egcs-2.91.66 with the -O2 optimisation flag.} is built up by
integrating the C++ code of the data part and the above C code.

\subsection{Performance Comparison}\label{perfanaly}
\begin{figure}[htbp]
\centerline{\includegraphics[width=8.5cm]{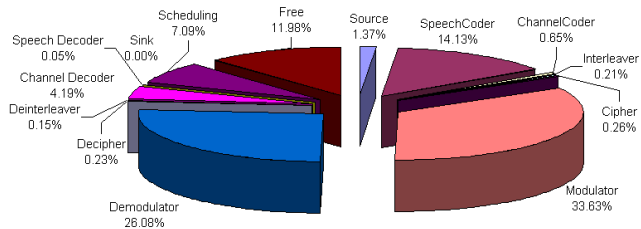}}
\caption{CPU requirement (\%) for 7 logical channels in EPspectra}
\label{cpu7ch_epspectra}
\end{figure}

\begin{figure}[htbp]
\centerline{\includegraphics[width=8.5cm]{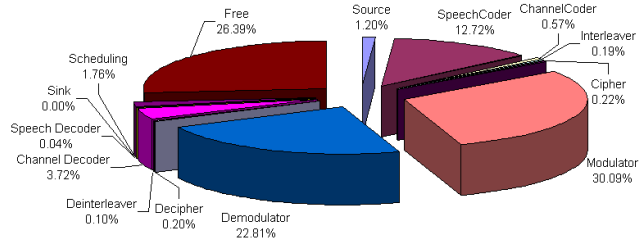}}
\caption{CPU requirement (\%) for 7 logical channels in Pspectra}
\label{cpu7ch_pspectra}
\end{figure}

We provide the performance comparison of GSM applications built on
{\sc EPspectra} and {\sc Pspectra}. Figure~\ref{cpu7ch_pspectra} and
Figure~\ref{cpu7ch_epspectra} show the CPU requirement for the GSM
programs to operate seven logical channels, respectively. Each
logical channel includes the operations of the transmission from
source information to radio waves and back.
With respect to the scheduling segment, the GSM program implemented
in {\sc EPspectra} consumes CPU four times more than in {\sc
Pspectra}. It is because of the interface part which provides no
scheduling functionality but connection between {\sc Esterel} code
and C++ code. Partially, the scheduling performance also varies to a
large extent of the optimization of automatically generated code
from {\sc Esterel}. However, this overhead has no effect on the
channel handling capability of signal processing process. For
example, in Figure~\ref{cpu7ch_epspectra} 11.98 percent of CPU are
still available as free. Therefore, the performance in terms of the
number of handled channels is the same.

\subsection{Comparison of LoC (Lines of Code)}
\begin{table}[hptb]
\caption{Comparison of loc}
\label{cmp-line}
\begin{tabular}{c|c|c}
\hline
code \textbackslash{} model &{\sc EPspectra} &{\sc Pspectra}\\
\hline
\hline
{\sc Esterel} code  & 1434& {\it not used} \\
\hline
{\sc Esterel}-generated C code  & 10876& {\it not used} \\
\hline
C/C++ code for interface part   & 3181& {\it not used} \\
\hline
C++ code for control part   & {\it not used}& 8045 \\
\hline
\hline
Assembly code   & 64310& 13024\\
\hline
\end{tabular}
\end{table}

Table \ref{cmp-line} makes the comparison of the loc of the control
part of {\sc EPspectra} and {\sc Pspectra} for the GSM
programs\footnote{The code corresponding to the data part of {\sc
EPspectra} is the same as is used in {\sc Pspectra}.}. The control
part of {\sc EPspectra} contains 1434 lines of the {\sc Esterel}
code and the {\sc Esterel} code is translated into the C code
corresponding to 10876 lines. The code used for the interface
contains the C code for the access to the control part and the C++
code for the access to the data part.

With comparison of the assembly code composed of only the executable
code, the loc corresponding to {\sc EPspectra} is 4.9 times larger
than the loc corresponding to {\sc Pspectra}. Nevertheless, given
that the advantage of a general purpose system is to use the large
amount of memory, the loc is not an important issue for these
applications, as opposed to embedded applications. Instead, the cost
of extra loc can be absorbed by the benefit of the {\sc Esterel}
methodology: simulation and verification.\\

{\bf Difficulties}: Programmers with {\sc EPspectra} need to be
familiar with programming in {\sc Esterel}. In addition, in terms of
a degrading performance, this may be a fundamental constraint that
results from automatically generated codes.
It needs efficient techniques such as innovative scheduling techniques.\\

{\bf Advantages}: {\sc EPspectra}, whose features include the
simulation and verification phases, facilitates the design and
implementation of DSP applications. Moreover, it allows one to
directly verify the actual code of {\sc Esterel} programs that are
compiled into an executable code. It guarantees that the {\sc
Esterel} programs satisfy the safety properties so long as all
source code is proved correct and compiled to the targeted code.

\section{Related Work}\label{rel}
\cite{hal:92,hal:93} presented an example of specifying and
verifying a real-time program using a synchronous data-flow
language, {\sc Lustre}. They introduced a subway control system
which operates in a U-turn section. First, the subway control
system, in which two verifiable problems, collision and derailment
may happen, is specified in {\sc Lustre}. Next, the critical
properties are expressed as the invariance of some boolean {\sc
Lustre} expression. Temporal properties are handled with the
allowance of references to the past with respect to the current
instant. Once the environment representing behaviors of the subway
control system and its properties to be verified are done in {\sc
Lustre}, they are verified whether the assertions are true or false
using {\sc Lesar}, its associated verification tool. The
verification process runs relying on 'standard' model
checking~\cite{clarke:86} which leads to explicitly enumerating the
reachable states and symbolic model checking~\cite{burch:90} which
starts from a boolean formula and iteratively computes a sequence of
formulas. The advantage of the work is that there is no manual
transformation between the program that is verified and the code
that is executed.

\cite{hal:97} presented linear relation analysis applied to the
verification of quantitative time properties of both synchronous
programs and linear hybrid systems.

\cite{jeannet:99} proposed to dynamically select a suitable
partitioning according to the property to be proved, avoiding
exponential explosion of the analysis caused by in-depth detailed
partitioning.

\cite{raymond:98,hal:99} proposed to use synchronous observers to
express both the relevance and the correctness of the test
sequences. The relevance observer is used to randomly choose inputs
satisfying temporal assumptions about the environment.

\cite{guernic:92,borgne:96} presented an example of verification of
real-time applications, using a synchronous language, {\sc Signal}.
The overall procedure from programming to verification is similar to
that using {\sc Lustre}. {\sc Signal} approach provides the ease of
implementing distributed systems including the features of proof and
compilation.

\section{Discussion}
We have presented {\sc EPspectra} for DSP applications development
and verification. {\sc EPspectra} methodology achieves a substantial
principle of {\it what we prove is what we execute}~\cite{berry:89}
straightforwardly; there is no gap between the program which is
verified and the code which is executed. All specification,
simulation, verification and execution are performed in it. We have
also demonstrated the implementation and verification of the radio
interface part of a GSM BTS using {\sc EPspectra}. The performance
results are promising in that the benefit from the verification
functionality absorbs the impact on the overhead of automated
generated code.
\\

In future work, we shall experiment with the automatic
test-generation feature. The {\sc Esterel} model-checker {\sc Xeve}
provides an automatic test-cases generation feature that can further
reduce the time cost of the testing phase~\cite{arditi:99}: the
generated test-cases are such that the {\sc Esterel} FSM model's
states are totally covered, that is, every state of the model is
visited and stimulated at least once by the test cases. With these
test cases, the developers can detect more potential tricky bugs
called corner cases, which are particularly hard to write a test
case for.

We shall also attempt to verify timing constraints considering that
the applications developed by {\sc EPspectra} correspond to
time-sensitive systems based on either hard real-time constraints or
soft real-time constraints. The method introduced
in~\cite{closse:03} can be used to verify quantitative timing
constraints by using a time-driven automata.

\section*{Acknowledgements}
We gratefully acknowledge discussions about {\sc Pspectra} with John
C. Ankcorn at MIT and about {\sc Esterel} Verification Techniques
with Robert de Simone at INRIA and about Timing Constraints
Verification Techniques with Daniel Weil and Jacques Pulou at France
Telecom R\&D and about Software Development Methodology with Thierry
Saunier at Thales. This research was supported by the DESS project
associated with Information Technology for European Advancement
(ITEA).

%
%
\bibliography{./kim05epspectra}
\appendix
\section*{Appendix A. Glossary}
\begin{itemize}
\item [MS]  The GSM mobile station (or mobile phone) communicates
with other parts of the system through the base-station system.
\item [GSM] Global System for Mobile communications is the European
standard for digital cellular telephone service.
\item [BTS] The Base Transceiver Station handles the radio interface
to the mobile station. The base transceiver station is the radio
equipment (transceivers and antennas).
\item [BSS] GSM Base Station Subsystem provides the interface
between the GSM mobile phone and other parts of the GSM network. It
consists of one or more base transceiver station (BTS) and one or
more base station controller (BSC).
\item [NSS] Network SubSystem performs the switching of calls
between the mobile users, and between mobile and fixed network users.
\item [MSC] Mobile Switching Center performs the telephony
switching functions of the system. It also performs such functions
as toll ticketing, network interfacing, common channel signalling,
and others.
\item [BSC] Base Station Controller provides the control functions
and physical links between the MSC and BTS. It provides functions
such as handover, cell configuration data and control of RF power
levels in base transceiver stations.
\item [HLR] Home Location Register database is used for storage and
management of subscriptions. The home location register stores
permanent data about subscribers, including a subscriber's service
profile, location information, and activity status.
\item [VLR] Visitor Location Register database contains temporary
information about subscribers that is needed by the MSC in order to
service visiting subscribers.
\item [EIR] Equipment Identity Register database contains
information on the identity of mobile equipment to prevent calls
from stolen, unauthorized or defective mobile stations.
\item [AuC] Authentication Center provides authentication and
encryption parameters that verify the user's identity and ensure the
confidentiality of each call.
\item [DSP] Digital Signal Processing are specialized computer chips
designed to perform speedy and complex operations on digitized
waveforms. It is used in processing sound, such as voice phone
calls, and video.
\item [RPE/LTP] Regular Pulse Excitation with Long Term Prediction
is used by GSM for full rate speech coding.
\item [GMSK] Gaussian Minimum Shift Keying is the modulation technique
used in GSM networks. It employs a form of FSK (Frequency Shift
Keying).
\item [GuPPI] General Purpose PCI I/O is the PCI appliance base for
the SpectrumWare project. Its function is to provide an efficient
means for moving a continuous stream of sampled data between a
workstation's main memory and an application-specific analog
daughtercard.\\ See URL:
http://www.sds.lcs.mit.edu/SpectrumWare/guppi.html.
\item [QAM] Quadrature Amplitude Modulation is a method for encoding
digital data in an analog signal in which each combination of phase
and amplitude represents one of sixteen four bit patterns. This is
required for fax transmission at 9600 bits per second. This
constellation, and therefore the number of bits which can be
transmitted at once, can be increased for higher bit rates and
faster throughput, or decreased for more reliable transmission with
fewer bit errors. The number of "dots" in the constellation is given
as a number before the QAM, and is always two to the power of an
integer from one (2QAM) to twelve (4096QAM). 64QAM is often used in
digital cable television and cable modem applications.
\end{itemize}

\section*{Appendix B. Source Code}
The complete source code of {\sc EPspectra} is available in a public
domain for the purpose of research. See {\tt
http://www.inria.fr/planete/hkim/epspectra/}.  The GSM radio
interface implementation consists of the downlink and uplink part.
We present main {\sc Esterel} and C code of the downlink part,
respectively, in Appendix B.1 and B.2. The {\sc Esterel} code in
Appendix B.1 is the one that is verified and compiled/executed. Once
it is translated into the corresponding C code with {\sc Esterel}
compiler,  main function in Appendix B.2 calls the DNLINK function
originated from the {\sc Esterel} code. Each time DNLINK() is called
in main function, a logical unit that is identified by the statement
from a 'tick' to the next is executed in {\sc Esterel} code.
\subsection*{B.1 Downlink {\sc Esterel} code}
\begin{verbatim}
%##########################################################
%#  This module is downlink application with data flow model.
%##########################################################

module DNLINK:

type StrlSampleRange;
type UnsignedLL;
type UnsignedLong;

constant INITIAL_RANGE:StrlSampleRange;
constant INITIAL_UNSIGNEDLL:UnsignedLL;

%%%%%%%%%%%%%%%%%%%%%%%%%%%%
% parameter of modules
%%%%%%%%%%%%%%%%%%%%%%%%%%%%

constant RATE1 = 32000 : integer; %{160*50}%
constant RATE2 = 6600 : integer; %{33*50}%
constant RATE3 = 91200: integer; %{456*50}%
constant RATE4 = 118400: integer; %{592*50}%
constant RATE5 = 177600: integer; %{148*6*50}%

input on_TimeConstraint:integer;
input IP_Addr:string;
input User_Quit;
input InitRange:StrlSampleRange; %{0 1600}%
inputoutput FileSource_module:string;
inputoutput SpeechCoder_module:string;
inputoutput ChannelCoder_module:string;
inputoutput Interleaver_module:string;
inputoutput Cipher_module:string;
inputoutput Modulator_module:string;
inputoutput UDPSink_module:string;
function GET_FILESOURCE(string,integer):string;
function GET_SPEECHCODER():string;
function GET_CHANNELCODER():string;
function GET_INTERLEAVER():string;
function GET_CIPHER():string;
function GET_MODULATOR():string;
function GET_UDPSINK(string,integer):string;

procedure CONNECT_MODULES()(string,string,integer,integer);
procedure INITIAL_SINK()(string);
%
% body part
%
signal Mark_src2spcoder:=INITIAL_RANGE:StrlSampleRange,
       Mark_spcoder2chcoder:=INITIAL_RANGE:StrlSampleRange,
       Mark_chcoder2inleaver:=INITIAL_RANGE:StrlSampleRange,
       Mark_inleaver2cipher:=INITIAL_RANGE:StrlSampleRange,
       Mark_cipher2mod:=INITIAL_RANGE:StrlSampleRange,
       Mark_mod2snk:=INITIAL_RANGE:StrlSampleRange,
       Compute_src2spcoder, Compute_spcoder2chcoder,
       Compute_chcoder2inleaver, Compute_inleaver2cipher,
       Compute_cipher2mod, Compute_mod2snk,
       Ack_snk2mod:=INITIAL_RANGE:StrlSampleRange,
       Ack_mod2cipher:=INITIAL_RANGE:StrlSampleRange,
       Ack_cipher2inleaver:=INITIAL_RANGE:StrlSampleRange,
       Ack_inleaver2chcoder:=INITIAL_RANGE:StrlSampleRange,
       Ack_chcoder2spcoder:=INITIAL_RANGE:StrlSampleRange,
       Ack_spcoder2src:=INITIAL_RANGE:StrlSampleRange,
       RDV_snk2mod, RDV_mod2cipher, RDV_cipher2inleaver,
       RDV_inleaver2chcoder, RDV_chcoder2spcoder,
       RDV_spcoder2src,Ready2Receive
in
%%%%%%%%%%%%%%%%%%%%%
% create modules
%%%%%%%%%%%%%%%%%%%%%
abort
   await IP_Addr;
   emit FileSource_module(GET_FILESOURCE("papin2.au",0));
   emit SpeechCoder_module(GET_SPEECHCODER());
   emit ChannelCoder_module(GET_CHANNELCODER());
   emit Interleaver_module(GET_INTERLEAVER());
   emit Cipher_module(GET_CIPHER());
   emit Modulator_module(GET_MODULATOR());
   emit UDPSink_module(GET_UDPSINK(?IP_Addr,5001));
%%%%%%%%%%%%%%%%%%%%%%%%%%
% make topology
%%%%%%%%%%%%%%%%%%%%%%%%%%
  call CONNECT_MODULES()(?UDPSink_module,?Modulator_module,RATE5,8);
  call CONNECT_MODULES()(?Modulator_module,?Cipher_module,RATE4,8);
  call CONNECT_MODULES()(?Cipher_module,?Interleaver_module,RATE4,8);
  call CONNECT_MODULES()(?Interleaver_module,?ChannelCoder_module,RATE3,8);
  call CONNECT_MODULES()(?ChannelCoder_module,?SpeechCoder_module,RATE2,8);
  call CONNECT_MODULES()(?SpeechCoder_module,?FileSource_module,RATE1,8);

  call INITIAL_SINK()(?UDPSink_module);
  await InitRange;
%%%%%%%%%%%%%%%%%%%%%%%%%%
% initialize parameters
%%%%%%%%%%%%%%%%%%%%%%%%%%
  [
     emit Ack_spcoder2src(?InitRange);
   ||
     run FileSource/
    SOURCE[signal FileSource_module/Name;
           signal Mark_src2spcoder/E_Mark_to_Down;%{mark1}%
           signal Compute_src2spcoder/E_Compute_to_Down;
           signal Ack_spcoder2src/Ack_From_Down;
           signal RDV_spcoder2src/snooping%{;
           signal FileSource_COMPUTEDSR/ComputedSRange}%];
   ||
      run SpeechCoder/
    P_MOD[signal SpeechCoder_module/Name;
          signal Mark_src2spcoder/S_Mark_from_Up;%{mark1}%
          signal Compute_src2spcoder/S_Compute_from_Up;%{}%
          signal Ack_spcoder2src/Ack_to_Up;%{}%
          signal RDV_spcoder2src/sig_on;%{}%
          signal Ready2Receive/Ready2Receive;
          signal Mark_spcoder2chcoder/E_Mark_to_Down;%{mark2}%
          signal Compute_spcoder2chcoder/E_Compute_to_Down;%{wire2}%
          signal Ack_chcoder2spcoder/Ack_From_Down;%{wire3}%
          signal RDV_chcoder2spcoder/snooping%{;
          signal SpeechCoder_COMPUTEDSR/ComputedSRange}%];
   ||
      run ChannelCoder/
    P_MOD[signal ChannelCoder_module/Name;
          signal Mark_spcoder2chcoder/S_Mark_from_Up;%{mark2}%
          signal Compute_spcoder2chcoder/S_Compute_from_Up;%{wire2}%
          signal Ack_chcoder2spcoder/Ack_to_Up;%{wire3}%
          signal RDV_chcoder2spcoder/sig_on;%{wire4}%
          signal Ready2Receive/Ready2Receive;
          signal Mark_chcoder2inleaver/E_Mark_to_Down;%{snk1}%
          signal Compute_chcoder2inleaver/E_Compute_to_Down;%{}%
          signal Ack_inleaver2chcoder/Ack_From_Down;%{from sink1}%
          signal RDV_inleaver2chcoder/snooping%{;
          signal ChannelCoder_COMPUTEDSR/ComputedSRange}%];
   ||
      run Interleaver/
    P_MOD[signal Interleaver_module/Name;
          signal Mark_chcoder2inleaver/S_Mark_from_Up;%{mark2}%
          signal Compute_chcoder2inleaver/S_Compute_from_Up;%{wire2}%
          signal Ack_inleaver2chcoder/Ack_to_Up;%{wire3}%
          signal RDV_inleaver2chcoder/sig_on;%{wire4}%
          signal Ready2Receive/Ready2Receive;
          signal Mark_inleaver2cipher/E_Mark_to_Down;%{snk1}%
          signal Compute_inleaver2cipher/E_Compute_to_Down;%{}%
          signal Ack_cipher2inleaver/Ack_From_Down;%{from sink1}%
          signal RDV_cipher2inleaver/snooping%{;
          signal Interleaver_COMPUTEDSR/ComputedSRange}%];
   ||
      run Cipher/
    P_MOD[signal Cipher_module/Name;
          signal Mark_inleaver2cipher/S_Mark_from_Up;%{mark2}%
          signal Compute_inleaver2cipher/S_Compute_from_Up;%{wire2}%
          signal Ack_cipher2inleaver/Ack_to_Up;%{wire3}%
          signal RDV_cipher2inleaver/sig_on;%{wire4}%
          signal Ready2Receive/Ready2Receive;
          signal Mark_cipher2mod/E_Mark_to_Down;%{snk1}%
          signal Compute_cipher2mod/E_Compute_to_Down;%{}%
          signal Ack_mod2cipher/Ack_From_Down;%{from sink1}%
          signal RDV_mod2cipher/snooping%{;
          signal Cipher_COMPUTEDSR/ComputedSRange}%];
   ||
      run Modulator/
    P_MOD[signal Modulator_module/Name;
          signal Mark_cipher2mod/S_Mark_from_Up;%{mark2}%
          signal Compute_cipher2mod/S_Compute_from_Up;%{wire2}%
          signal Ack_mod2cipher/Ack_to_Up;%{wire3}%
          signal RDV_mod2cipher/sig_on;%{wire4}%
          signal Ready2Receive/Ready2Receive;
          signal Mark_mod2snk/E_Mark_to_Down;%{snk1}%
          signal Compute_mod2snk/E_Compute_to_Down;%{}%
          signal Ack_snk2mod/Ack_From_Down;%{from sink1}%
          signal RDV_snk2mod/snooping%{;
          signal Modulator_COMPUTEDSR/ComputedSRange}%];
   ||
      run UDPSink/
    SINK[signal UDPSink_module/Name;
         signal Mark_mod2snk/S_Mark_from_Up;%{snk2}%
         signal Compute_mod2snk/S_Compute_from_Up;
         signal Ack_snk2mod/Ack_to_Up;
         signal RDV_snk2mod/sig_on;
         signal Ready2Receive/Ready2Receive%{;
         signal UDPSink_COMPUTEDSR/ComputedSRange}%];
   ]
when User_Quit
end signal
end module
\end{verbatim}

\subsection*{B.2 Downlink main C code}
\begin{verbatim}
#include <stdio.h>
#include <sys/time.h>
#include "GSM_DNLINK.h"

main(int argc,char** argv){
  char *addr=(char *)malloc(sizeof(char[16]));
  if (argc < 2)
    strcpy(addr,"localhost");
  else
    strcpy(addr,argv[1]);
  DNLINK();
  DNLINK_I_IP_Addr(addr);
  DNLINK();
  DNLINK_I_InitRange("0 1600");
    while(1)
    DNLINK();
}
\end{verbatim}
\end{document}